\documentclass[aps,prl,twocolumn,superscriptaddress,showpacs,floatfix]{revtex4-1}

\usepackage{times}
\usepackage{graphicx}
\usepackage{dcolumn}
\usepackage{bm}
\usepackage{color}
\usepackage{longtable}
\usepackage{booktabs}
\usepackage{amssymb}
\usepackage{textcomp}
\usepackage{upgreek} 

\begin{document}

\title{Raising the superconducting $\bm{T_\mathrm{c}}$ of gallium:\\in-situ characterization of the transformation of $\bm{\alpha}$-Ga into $\bm{\beta}$-Ga}

\author{D. Campanini}
\affiliation{Department of Physics, Stockholm University, SE-106 91 Stockholm, Sweden}

\author{Z. Diao}
\affiliation{Department of Physics, Stockholm University, SE-106 91 Stockholm, Sweden}
\affiliation{School of Information Technology, Halmstad University, Box 823, SE-301 18 Halmstad, Sweden}

\author{A. Rydh}\email{andreas.rydh@fysik.su.se}
\affiliation{Department of Physics, Stockholm University, SE-106 91 Stockholm, Sweden}

\date{\today}

\begin{abstract}
Gallium (Ga) displays several metastable phases. Superconductivity is strongly enhanced in the metastable $\mathrm{\beta}$-Ga with a critical temperature $T_\mathrm{c}= 6.04(5)\,\mathrm{K}$, while stable $\mathrm{\alpha}$-Ga has a much lower $T_\mathrm{c}<1.2\,\mathrm{K}$. Here we use a membrane-based nanocalorimeter to initiate the transition from $\mathrm{\alpha}$-Ga to  $\mathrm{\beta}$-Ga on demand, as well as study the specific heat of the two phases on one and the same sample. The in-situ transformation is initiated by bringing the temperature to about $10\,\mathrm{K}$ above the melting temperature of $\mathrm{\alpha}$-Ga. After such treatment, the liquid supercools down to $232\,\mathrm{K}$, where $\mathrm{\beta}$-Ga solidifies.  We find that $\mathrm{\beta}$-Ga is a strong-coupling type-I superconductor with $\Delta(0)/k_\mathrm{B}T_\mathrm{c} =2.00(5)$ and a Sommerfeld coefficient $\gamma_\mathrm{n} = 1.53(4)\,\mathrm{mJ/molK^2}$, 2.55 times higher than that in the $\alpha$ phase. The results allow a detailed comparison of fundamental thermodynamic properties between the two phases.

\end{abstract}

\pacs{74.25.Bt, 74.62.Bf}

\maketitle

\section{Introduction}

Gallium is a metal with peculiar properties and very rich phase diagrams, caused by an interplay between covalent and metallic bonding. Like water, Ga expands as it freezes. The stable solid phase in ambient conditions ($\alpha$-Ga) has a melting point at 302.9\,K~\cite{CRC2012} and is superconducting below 1.08\,K~\cite{Seidel1958,Gregory1966}. Several metastable phases with different physical properties can, however, be produced at ambient pressure by supercooling the melt. The most commonly obtained are the $\beta$~\cite{Bosio1967,Bosio1969}, $\gamma$~\cite{Bosio1967,Bosio1972} and $\delta$-Ga~\cite{Bosio1973}. Reports on six additional phases, obtained under particular conditions, are also found~\cite{Bosio1968,Heyding1973,Teske1999,Lee2010}. They are characterized by different crystal structures, in turn leading to, \emph{e.g.}, different melting temperatures and superconducting critical temperatures.

\begin{figure}[b]
	\includegraphics[width=\linewidth]{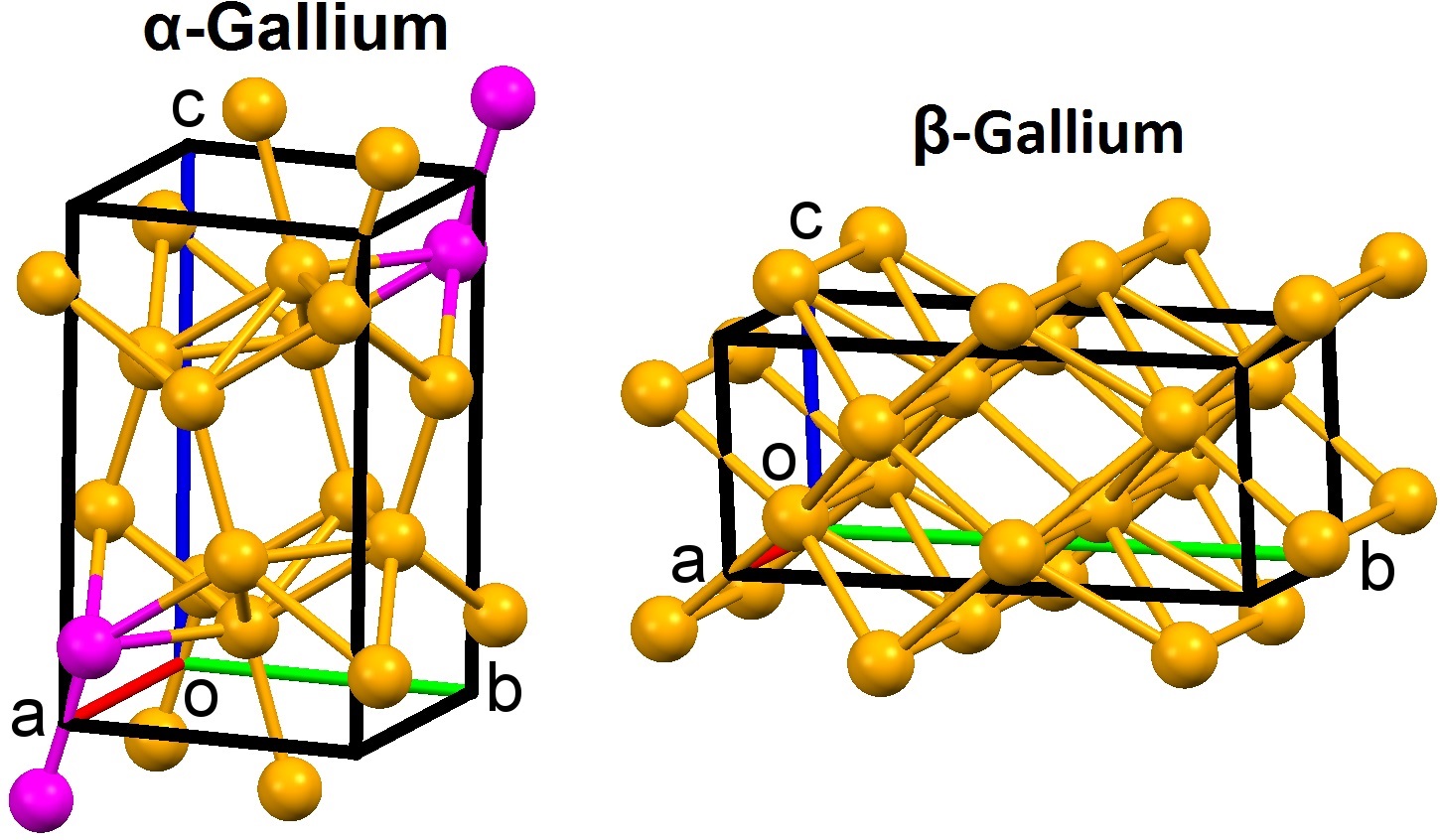}
	\caption{(a) Orthorhombic unit cell of $\alpha$-Ga with 8 atoms/cell. Two $\mathrm{Ga_2}$ dimers are represented in magenta. (b) Monoclinic unit cell of $\beta$-Ga with 4 atoms/cell.}
	\label{FigStructure}
\end{figure}

The crystal structures of $\alpha$ and $\beta$-Ga are given in Fig.~\ref{FigStructure}.
The stable $\alpha$-Ga has an orthorhombic unit cell~\cite{Wyckoff1963}, space group \emph{Cmca} $(D_\mathrm{2h}^{18})$, while $\beta$-Ga is monoclinic, space group $C2/c$ $(C_\mathrm{2h}^6)$~\cite{Bosio1969}. In $\alpha$-Ga metallicity is coexisting with covalently bonded $\mathrm{Ga_2}$ dimers between nearest neighbors~\cite{Gong1991}, while such dimers are absent in $\beta$-Ga. 

The metastable phases are typically obtained by dispersing small Ga droplets in a solvent~\cite{Bosio1967,Bosio1969,Parr1973,Heyding1973,He2005,Li2011}. Once cooled, a droplet can crystallize in one of the cited phases, with a probability to attain a certain phase depending on the droplet size~\cite{He2005,Li2011}. $\beta$-Ga is the most commonly produced as it presents the least difference in energy from the stable $\alpha$ phase~\cite{Bernasconi1995}. Spherical droplets with diameters up to 500\,$\mathrm{\upmu m}$ have been reported~\cite{Bosio1965}. Larger $\beta$-Ga crystals, with sizes of several mm, can be obtained under well-controlled conditions by crystallization of liquid Ga from a $\beta$-Ga seed~\cite{Bosio1981}. The samples are however easily relaxing to the $\alpha$ phase if exposed to shocks or vibrations. 

Due to the technical difficulties in handling metastable phases, relatively little is known about $\beta$-Ga in comparison with $\alpha$-Ga. Low-temperature specific heat data is so far not available. While it is clear that Ga is a type-I superconductor in both phases, experimental evidence is contradictory on whether $\beta$-Ga is a weak- or strong-coupling superconductor~\cite{Parr1974,Cohen1967,Wuhl1968}. In this work, we obtain $\beta$-Ga from $\alpha$-Ga samples with sizes up to 150\,$\mathrm{\upmu m}$ and masses up to $100\,\mathrm{\upmu g}$ by supercooling liquid Ga on top of a membrane-based nanocalorimeter in vacuum. The specific heat of $\alpha$- and $\beta$-Ga is measured from 315\,K down to 1.2\,K before and after the in-situ recrystallization into the metastable phase. This allows an accurate comparison of the specific heat and entropy of the two phases. 
From the low-temperature specific heat measurements we determine the Sommerfeld coefficient $\gamma_\mathrm{n}$,
the specific heat jump at the critical temperature $\Delta C/T_\mathrm{c}$, the superconducting condensation energy $\Delta F(0)$, the temperature dependence of the thermodynamic critical field $H_\mathrm{c}(T)$, and related quantities.
We conclude that $\beta$-Ga is a strong-coupling type-I superconductor with an increased density of states and a phonon system with increased entropy.

\section{Experimental details}

Ga samples with sizes between $10\,\mathrm{\upmu m}$ and $150\,\mathrm{\upmu m}$ were prepared and individually attached to a membrane-based ac-nanocalorimeter~\cite{Tagliati2011, Tagliati2012} using a small amount of Apiezon-N grease, see inset of Fig.~\ref{FigThermalCycle}(a). The measurements were performed with a fixed-phase variable-frequency method under vacuum conditions to obtain good absolute accuracy. A thin-film heater integrated on the device was used to locally control the sample temperature, which was measured with a thin-film Ge$_{1-x}$Au$_{x}$ thermometer in direct thermal contact with the sample~\cite{Tagliati2012}. The transition into $\beta$-Ga was initiated by heating the sample to several kelvin above the melting transition at $T_\mathrm{m,\alpha} = 302.9\,\mathrm{K}$. After bringing the temperature back down to below $T_\mathrm{m,\alpha}$, supercooling was observed down to either $285\,\mathrm{K}$, where solidification into the $\alpha$ phase takes place, or all the way down to $232\,\mathrm{K}$, with subsequent solidification into $\beta$-Ga, if the liquid was initially heated to a high enough temperature. At low temperatures, the specific heat was measured in magnetic fields up to 100\,mT to probe the superconducting properties of $\beta$-Ga.

\section{Results and discussion}
\begin{figure}[t]
	\includegraphics[width=\linewidth]{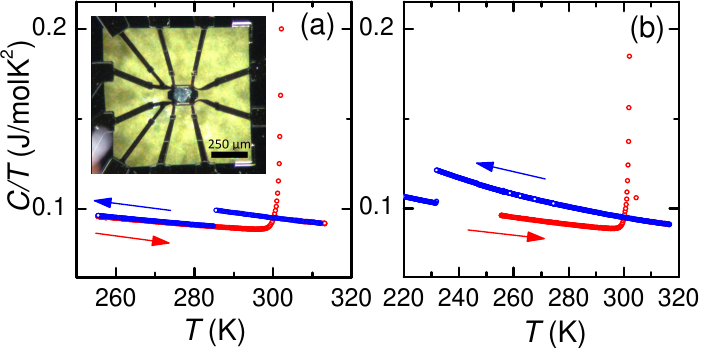}
	\caption{Temperature dependence of the specific heat of Ga, shown as $C/T$, measured on heating and cooling. (a) Heating from 255\,K up to 313\,K followed by a cooling back to 255\,K. The sample melts at 302.9\,K and solidifies to the $\alpha$ phase at 285\,K. Inset: optical microscope image of a Ga sample mounted onto the nanocalorimeter. (b) Heating from 255\,K up to 317\,K followed by a cooling down to 220\,K. The same sample melts at 302.9\,K and solidifies to the $\beta$ phase at 232\,K.} 
	\label{FigThermalCycle}
\end{figure}

The specific heat signatures measured during two heating and cooling cycles around the melting temperature $T_\mathrm{m,\alpha}$ are shown in Fig.~\ref{FigThermalCycle}.
In Fig.~\ref{FigThermalCycle}(a) the sample is heated up to $313\,\mathrm{K}$. Once cooled, it solidifies back to the $\alpha$ phase at $285\,\mathrm{K}$. When the same sample is instead heated up to $317\,\mathrm{K}$, it supercools all the way down to $232\,\mathrm{K}$, see Fig.~\ref{FigThermalCycle}(b). If reheated again, the sample melts at the melting temperature $T_\mathrm{m,\beta}=257\,\mathrm{K}$, a signature of $\beta$-Ga~\cite{Bosio1965,Bosio1967,Bosio1968,Heyding1973,Parr1973}. 

\begin{figure}[t]
	\includegraphics[width=\linewidth]{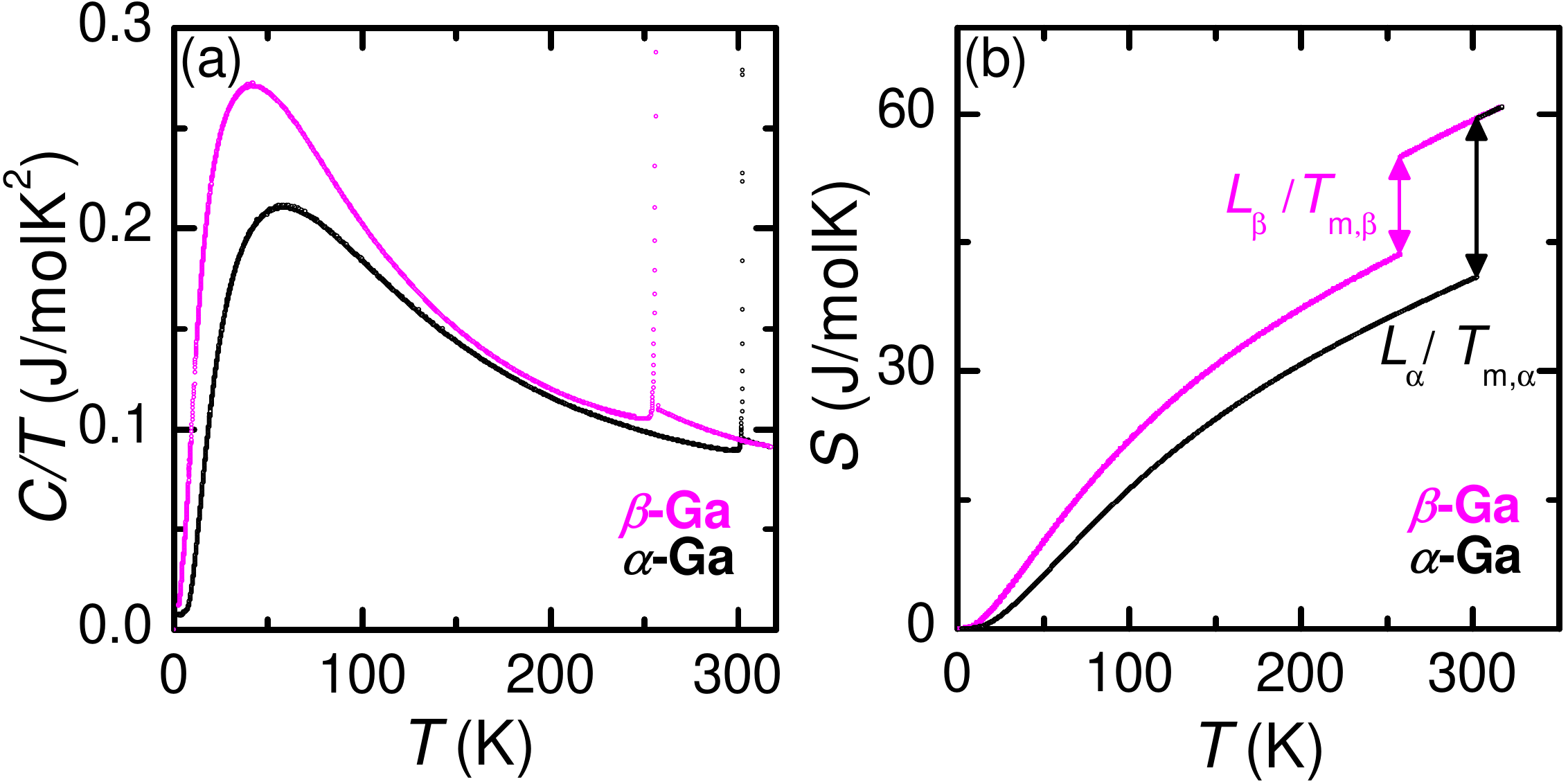}
	\caption{(a) Temperature dependence of the specific heat, shown as $C/T$, over the full temperature range. (b) Temperature dependence of the entropy.
}
	\label{FigCandS}
\end{figure}

The temperature dependence of the specific heat, shown as $C/T$, as well as the entropy $S(T) = \int_0^T (C/T^\prime) \mathrm{d}T^\prime$ for $\alpha$- and $\beta$-Ga in the studied temperature range are shown in Fig.~\ref{FigCandS}.
The contribution from the grease and the device specific heat has been independently measured and then subtracted. Melting can be seen as two sharp peaks in $C/T$ at $257\,\mathrm{K}$ and $302.9\,\mathrm{K}$ for $\beta$- and $\alpha$-Ga, respectively, due to the presence of latent heat. The entropy of the metastable $\beta$ phase lies above the stable $\alpha$ phase over the full temperature range. The sharp steps at the melting temperatures in the entropy curves are due to latent heat absorption. Latent heat for $\alpha$-Ga at the melting point is taken as $L_\mathrm{\alpha} = T_\mathrm{m,\alpha}\cdot\Delta S (T_\mathrm{m,\alpha}) = 5585\,\mathrm{J/mol}$~\cite{Adams1952}. In order for the entropy to be conserved in the common liquid state, the latent heat for $\beta$-Ga at the melting point has to be $L_\mathrm{\beta} = T_\mathrm{m,\beta}\cdot\Delta S (T_\mathrm{m,\beta})=2891\,\mathrm{J/mol}$. This value is about 10\% higher than that reported in Ref.~\onlinecite{Defrain1960}.

\begin{figure}[t]
	\includegraphics[width=\linewidth]{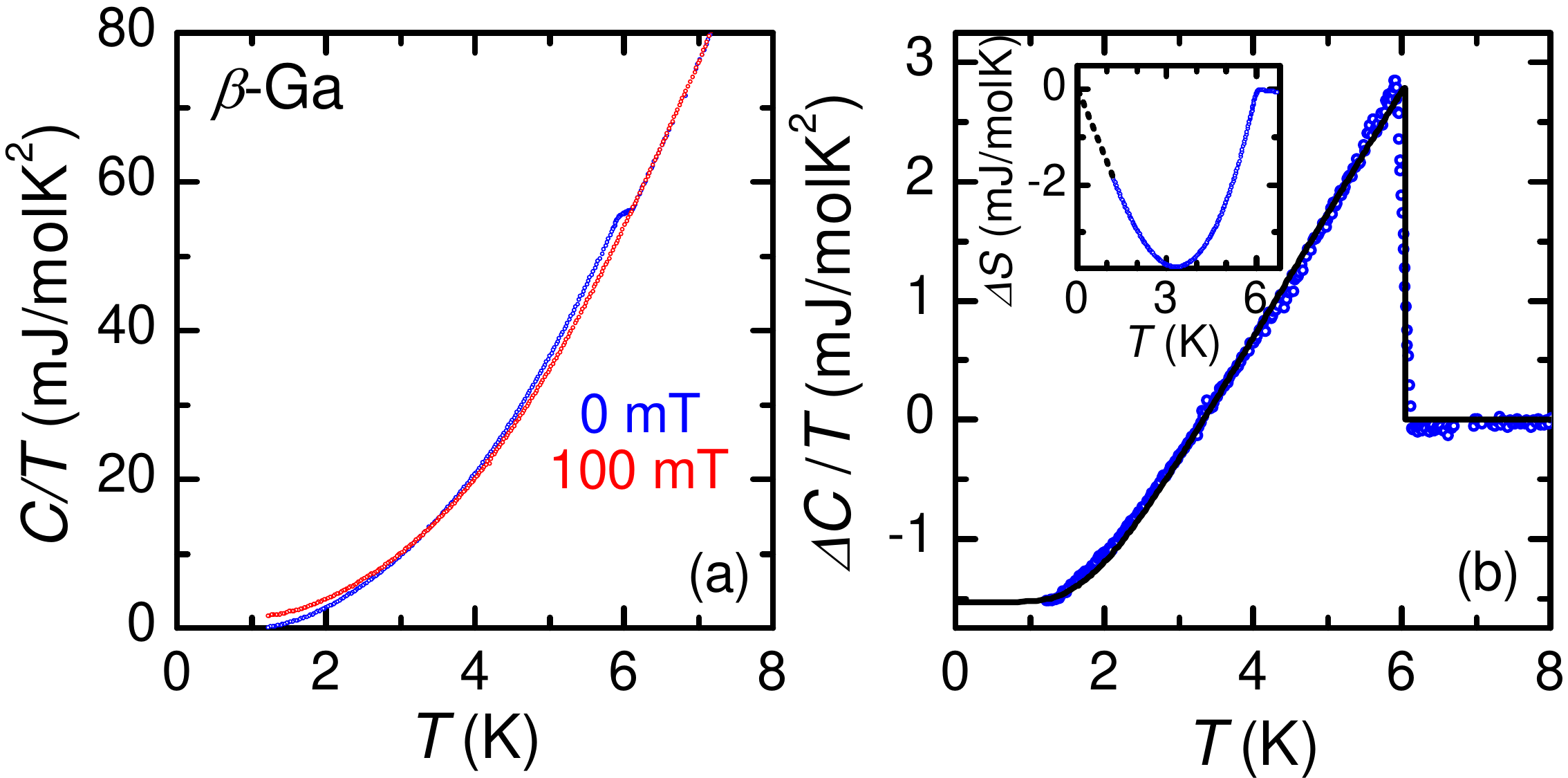}
	\caption{(a) Temperature dependence of the specific heat as $C/T$ for $\beta$-Ga at low temperatures in zero applied magnetic field (blue points) and for a field of 100\,mT (red points). (b) Temperature dependence of $\Delta C / T = [C(0\,\mathrm{mT}) - C(100\,\mathrm{mT})]/T$. Inset: Temperature dependence of the entropy difference $\Delta S$ between the superconducting and the normal state. Dashed lines are extrapolations to zero temperature.}
	\label{FigCLowT}
\end{figure}

The temperature dependence of the low temperature specific heat for $\beta$-Ga is shown in Fig.~\ref{FigCLowT}(a) as $C/T$ in zero field and with 100\,mT applied magnetic field. The step associated with the superconducting transition can be observed in the zero field curve at $T_\mathrm{c} = 6.04(5)\,\mathrm{K}$. No transition is observed for an applied field of $100\,\mathrm{mT}$. The 100\,mT curve is therefore taken as the normal state curve. The difference $\Delta C/T = [C(0\,\mathrm{mT})-C(100\,\mathrm{mT})]/T = (C_\mathrm{es} - C_\mathrm{en})/T $, where $C_\mathrm{es}$ indicates the electronic specific heat in the superconducting state and $C_\mathrm{en}$ in the normal state, is reported in Fig.~\ref{FigCLowT}(b).

In the inset of Fig.~\ref{FigCLowT}(b), the entropy difference  $\Delta S(T) = \int_0^T (\Delta C / T' )\mathrm{d}T'$ is shown, setting $\Delta S (T_\mathrm{c}) = 0$. Extrapolating to $\Delta S(0)=0$ leads to the low-temperature value $\Delta C/T|_{T\rightarrow 0}=-1.53\,\mathrm{mJ/molK^2}$, shown in the main panel of Fig.~\ref{FigCLowT}(b). The Sommerfeld term for $\beta$-Ga is thus $\gamma_\mathrm{n}=1.53(4)\,\mathrm{mJ/molK^2}$,
close to estimations from critical field curves of Ref.~\onlinecite{Parr1974} and around 2.55 times higher than that for $\alpha$-Ga~\cite{Phillips1964,Seidel1958}. Obtaining $\gamma_\mathrm{n}$ in this way avoids the effects of the contribution from the device and grease addenda heat capacity. However, there is a possibility that $\gamma_\mathrm{n}$ is underestimated if disorder or other effects make $C_\mathrm{es}/T|_{T\rightarrow 0} \ne 0$.

We could not obtain the corresponding superconducting properties or accurately estimate $\gamma_\mathrm{n}$ of $\alpha$-Ga with our current experimental setup because of the low $T_\mathrm{c}$ of $\alpha$-Ga. Values from the literature are, however, well known in this case.

From Fig.~\ref{FigCLowT}(b), the specific heat jump at $T_\mathrm{c}$ can be directly inferred: $\Delta C/T_\mathrm{c} = 2.80(5)\,\mathrm{mJ/molK^2}$. The value $\Delta C/\gamma_\mathrm{n}T_\mathrm{c}=1.83(7)$ is substantially higher than the 1.43 typical of a BCS weak-coupling superconductor. This indicates that the ratio $\alpha=\Delta(0)/k_\mathrm{B}T_\mathrm{c}$, where $\Delta(0)$ is the superconducting energy gap at zero temperature, should also be higher than the BCS value $\alpha_\mathrm{BCS}=1.76$. To verify this, the electronic specific heat $C_\mathrm{es}$ is fitted according to a single-gap s-wave $\alpha$ model~\cite{Padamsee1973,Johnston2013}. In this empirical model, $C_\mathrm{es}$ is described by a BCS-like description where $\alpha$ is not fixed to $\alpha_\mathrm{BCS}$ but is an adjustable parameter. A deviation from the experimental data is calculated for each $\alpha$ and the least deviation fit is reported in Fig.~\ref{FigCLowT}(b). The resulting parameter $\alpha = 2.00(5)$ is well above $\alpha_\mathrm{BCS}$ as expected. In absolute terms this corresponds to $\Delta(0) = 1.04(3)\,\mathrm{meV}$, in good agreement with tunneling measurements~\cite{Cohen1967,Wuhl1968} on thin films.

\begin{figure}[t]
	\includegraphics[width=\linewidth]{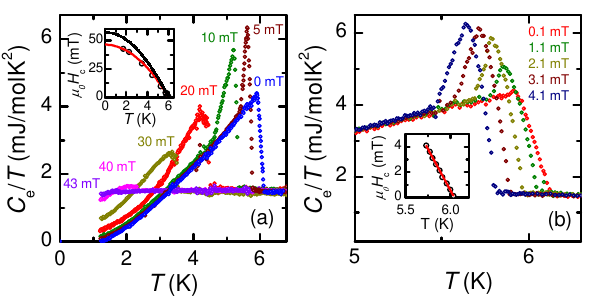}
	\caption{(a) Temperature dependence of the electronic specific heat as $C_\mathrm{e}/T$ for applied magnetic fields between 0 and 43\,mT. Inset: Temperature dependence of the critical field $H_\mathrm{c}$. The hollow circles correspond to the critical fields measured from the in-field curves. The red curve is a parabolic fit according to $H_\mathrm{c}(T)=H_\mathrm{c}(0)\left[1-\left(\frac{T}{T_\mathrm{c}}\right)^2\right]$. The black curve corresponds to $H_\mathrm{c}(T)$ calculated from $\Delta F(T)$. (b) Temperature dependence of the electronic specific heat as $C_\mathrm{e}/T$ for applied magnetic fields between 0.1 and 4.1\,mT. Inset: Temperature dependence of the critical field $H_\mathrm{c}$ near $T_\mathrm{c}$. The black circles correspond to the critical temperatures measured from the in-field curves. The red curve is a linear fit.}
	\label{FigMagnetic}
\end{figure}

By integrating $\Delta C$ and $\Delta C/T$ down from $T_\mathrm{c}$, the free energy difference between the superconducting and the normal state $\Delta F(T)$ is obtained. Using the $\alpha$-fit to extrapolate down to $T=0$, we obtain $-\Delta F(0) = 14.4(4)\,\mathrm{mJ/mol}$. 
The superconducting energy gap $\Delta(0)$ can be estimated from the free energy $\Delta F(0)$ using the BCS relation:
\begin{equation}
-\Delta F(0) = \frac{1}{4}N(E_\mathrm{F})\Delta^2(0).
\label{Eq:CondEnergy}
\end{equation}
From $\gamma_\mathrm{n} = \frac{\pi^2}{3}k_\mathrm{B}^2N(E_\mathrm{F})$, the density of states at the Fermi energy $N(E_\mathrm{F})$ can be estimated as $0.65(2)\,\mathrm{states}/\mathrm{eV}\/\mathrm{atom}$. A strong increase of $N(E_\mathrm{F})$ in comparison with $\alpha$-Ga is in agreement with both theoretical predictions~\cite{Bernasconi1995} and Knight shift data~\cite{Stroud1975}. Inserting this value into Eq.~(\ref{Eq:CondEnergy}) gives $\Delta(0)/k_\mathrm{B}T_\mathrm{c}=1.83(3)$, which corresponds to a $9\%$ underestimation of the value found through the $\alpha$ model. The difference can be attributed to a lower $\Delta F$ for strong-coupling superconductors than what expected in the weak-coupling limit~\cite{Haslinger2003,Carbotte1990}.
From $\Delta F(T)$ one can as well obtain the thermodynamic critical field $H_\mathrm{c}(T)$ through the definition $-\Delta F/V_\mathrm{m} = \mu_{0}H^2_\mathrm{c}/2$, where $V_\mathrm{m}$ is the molar volume. The resulting $H_\mathrm{c}(T)$ dependence is shown in the inset of Fig.~\ref{FigMagnetic}(a) as a black curve.
The zero temperature critical field is estimated to be $\mu_0H_\mathrm{c}(0)=57(2)\,\mathrm{mT}$, while the slope close to $T_\mathrm{c}$ is found to be $\left|{\mathrm{d}H^*_\mathrm{c}/\mathrm{d}T}\right|_{T_\mathrm{c}} = 15(1)\,\mathrm{mT/K}$.

It is possible to experimentally verify $H_\mathrm{c}(T)$ by studying the specific heat in applied magnetic fields. The temperature dependence of $C_\mathrm{e}/T$ for applied fields between 0 and 43\,mT is shown in Fig.~\ref{FigMagnetic}(a), with low-field curves near the superconducting transition shown in  Fig.~\ref{FigMagnetic}(b). The location of the transition, taken as the middle of the step for the different applied fields, is reported in the insets of Figs.~\ref{FigMagnetic}(a) and (b) as black circles. As seen, the directly measured transition, with an estimated $\mu_0H_\mathrm{c}(0)=46.5(1)\,\mathrm{mT}$, and $\left|{\mathrm{d}H^*_\mathrm{c}/\mathrm{d}T}\right|_{T_\mathrm{c}} = 13.1(2)\,\mathrm{mT/K}$, are about $15$-$20\%$ lower in comparison to $H_\mathrm{c}(T)$ evaluated thermodynamically from $\Delta F(T)$. This discrepancy is likely to be attributed to the demagnetization factor of the sample.

The sharp, additional peak in the vicinity of the transition seen in Fig.~\ref{FigMagnetic} is due to the latent heat of the transition, which is first order in the presence of a magnetic field. By integrating the excess specific heat peak, it is possible to estimate the amount of latent heat sensed by the calorimeter. For all fields up to 20\,mT, it corresponds to about half of the value calculated from $\Delta S$ of Fig.~\ref{FigCLowT}(b). For a thin slab in parallel with the magnetic field, one would expect all the latent heat to be released at $H_\mathrm{c}(T)$. However, due to the demagnetization factor of the sample, some areas become normal before others and the system is in a two-phase intermediate state over a certain temperature range, extending all the way down to $T=0$ for the highest fields. This makes the determination of $H_\mathrm{c}(T)$ somewhat ambiguous. The strongest demagnetization effect occurs at the low-temperature side of the latent heat peak, where full Meissner screening is in place. We take $H_\mathrm{c}(T)$ at the middle of the step, which remains fairly well-defined at all fields.

The parameters obtained from our measurement for $\beta$-Ga are summarized in Table~\ref{tab:SCParameters}. A comparison with $\alpha$-Ga is given as well, using available data from the literature.
The density of states $N(E_\mathrm{F})$ obtained from specific heat through the Sommerfeld coefficient represents the dressed/renormalized value. Band-structure calculations, on the other hand, give the bare density of states $N_{0}(E_\mathrm{F})$. For $\beta$-Ga,  $N_{0}(E_\mathrm{F}) = 0.42\,\mathrm{states}/\mathrm{eV}\/\mathrm{atom}$ agrees well with a simple free-electron model \cite{Stroud1975}. For $\alpha$-Ga, $N_{0}(E_\mathrm{F})$  is strongly reduced in comparison to such a model because of a pronounced pseudogap at the Fermi energy~\cite{Reed1969,Hunderi1974, Gong1991, Bernasconi1995}. It can be estimated as $0.12\,\mathrm{states}/\mathrm{eV}\/\mathrm{atom}$ if considered to be approximately 3.5 times smaller than that for $\beta$-Ga, as suggested by band structure calculations~\cite{Bernasconi1995}. Both $N(E_\mathrm{F})$ values calculated in this way are lower than the experimental values obtained from specific heat. If we assume that the difference is coming from an effective mass enhancement $m^*/m$ due to electron-phonon coupling, the electron-boson renormalization parameter $\lambda$ can be estimated from $m^*/m = 1 + \lambda=N(E_\mathrm{F})/N_{0}(E_\mathrm{F})$. This gives a $\lambda_\mathrm{\alpha} = 1.08$ for $\alpha$-Ga and $\lambda_\mathrm{\beta} = 0.55$ for $\beta$-Ga.

In a renormalized BCS frame, $T_\mathrm{c}$ is given by three parameters~\cite{Grimvall,Nakajima}:
\begin{equation}
T_\mathrm{c} = 1.134\cdot\theta_\mathrm{D}\exp\left[-\frac{1+\lambda}{\lambda - \mu^*(1+\lambda)}\right].
\label{Eq:BCS}
\end{equation}
Here, $\theta_\mathrm{D}$ is the Debye temperature and $\mu^*$ is the Coulomb pseudopotential coming from electron-electron interactions and opposing superconductivity.
The Debye temperature $\theta_\mathrm{D}$ for $\alpha$- and $\beta$-Ga can be evaluated by fitting $C/T$ as a function of $T^2$ at low temperatures.
We obtain $\theta_\mathrm{D,\alpha} = 350(20)\,$K and $\theta_\mathrm{D,\beta} = 120(20)\,$K. We note that the apparent $\theta_\mathrm{D}$ for both $\alpha$- and $\beta$-Ga are temperature dependent, but with opposite trends. For $\alpha$-Ga the trend agrees with the observations in Ref.~\onlinecite{Phillips1964}.
Inserting $T_\mathrm{c}$, $\theta_\mathrm{D}$, and $\lambda$ into Eq.~(\ref{Eq:BCS}), we find $\mu^*_\alpha = 0.35$ and $\mu^*_\beta = 0.03$ for $\alpha$- and $\beta$-Ga, respectively. Using McMillan's formula \cite{McMillan1968} results in $\mu^*_\alpha = 0.41$ and $\mu^*_\beta = -0.05$.
While the electron-boson renormalization parameter $\lambda$ is comparable in the two phases, the density of states $N(E_\mathrm{F})$ is much higher in $\beta$-Ga at the same time as the Coulomb pseudopotential $\mu^*$ is reduced from a fairly high value to nearly zero. These changes in combination explain the enhanced $T_\mathrm{c}$ in the metastable $\beta$ phase, or possibly more accurately the reduced $T_\mathrm{c}$ in $\alpha$-Ga.

\begin{table}[t]
\caption{Superconductivity-related parameters for $\alpha$- and $\beta$-Ga.}
\begin{tabular*}{\linewidth}{@{\extracolsep{\fill}}lccc}
	\hline
	\hline
	Property  & Unit & $\alpha$-Ga & $\beta$-Ga\\
	\hline\\[-2.2ex]
	$T_\mathrm{c}$ & K & \footnote{From Refs.\,\onlinecite{Seidel1958,Phillips1964,Gregory1966}.}1.08(1) & 6.04(5)\\
	$\gamma_\mathrm{n}$ & $\mathrm{mJ/molK^2}$ & \footnote{From Refs.\,\onlinecite{Seidel1958,Phillips1964}.}0.600(5) & 1.53(4)\\
	$\Delta C/\gamma_\mathrm{n}T_\mathrm{c}$ & & \footnote{From Ref.\,\onlinecite{Seidel1958}.}1.41 & 1.83(7) \\
	\footnote{Calculated from $\gamma_\mathrm{n}$ for $\alpha$-Ga using the relation $\gamma_\mathrm{n} = \frac{\pi^2}{3}k_\mathrm{B}^2N(E_\mathrm{F})$.}$N(E_\mathrm{F})$ & $\mathrm{states}/\mathrm{eV}\/\mathrm{atom}$ &0.25(1) & 0.65(2)\\
	$-\Delta F(0)$ & mJ/mol & \footnote{From integration of $\Delta C$ data taken from Ref.~\onlinecite{Seidel1958}.}0.17(1) & 14.4(4)\\
	$\mu_{0}H_\mathrm{c}(0)$ & mT &\footnote{From Ref.\,\onlinecite{Goodman1951}.} 5.03 & \footnote{From $\Delta F(0).$}57(2)\\
	$\Delta(0)/k_\mathrm{B}T_\mathrm{c}$ &  &\footnote{Value strongly dependent on the crystallographic axis, see Ref.\,\onlinecite{Morisseau1976} and references therein.}1.65-1.97  & 2.00(5)\\
	$\Delta(0)$ & meV & \footnote{calculated from $\Delta(0)/k_\mathrm{B}T_\mathrm{c}$ taken from Ref.\,\onlinecite{Morisseau1976}.}0.15-0.18 & 1.04(3).\\
	\footnote{From bare band-structure calculations.}$N_{0}(E_\mathrm{F})$& $\mathrm{states}/\mathrm{eV}\/\mathrm{atom}$ &\footnote{From Ref.~\onlinecite{Bernasconi1995}.}0.12 & \footnote{From Ref.~\onlinecite{Stroud1975}.}0.42\\
	$\lambda$ & & 1.08 & 0.55 \\
	\footnote{From low-temperature $C(T)$.}$\theta_\mathrm{D}$ & K & 350(20) & 120(20)\\
	\footnote{From Eq.~(\ref{Eq:BCS}).}$ \mu^*$ & & 0.35 & 0.03 \\ 
	\hline
	\hline
\end{tabular*}
\label{tab:SCParameters}
\end{table}

\section{Conclusions}
We show that it is possible to obtain metastable $\beta$-Ga through a simple thermal cycle of $\alpha$-Ga around its melting temperature by using a membrane-based nanocalorimeter. This in-situ transformation enables the measurement of specific heat on the same Ga sample in the two different phases. We find that superconductivity is enhanced in the metastable phase, with $T_\mathrm{c}$ shifting up to 6.04(5)\,K, mainly because of a strong increase in density of states at the Fermi energy $N(E_\mathrm{F})$ and reduction in Coulomb pseudopotential $\mu^*$. From the specific heat jump  at $T_\mathrm{c}$ and the temperature dependence of $\Delta C(T)$, we estimate a zero temperature energy gap $\Delta(0) = 1.04(3)\,\mathrm{meV}$ and conclude that $\beta$-Ga is a strong-coupling superconductor
.

\section{Acknowledgments}
This work was supported by the Swedish Research Council under grant number 2016-04516 (A.R.) and grant number 2015-00585 (Z.D.), the latter co-funded by Marie Sklodowska-Curie Actions (Project INCA 600398). D.C. acknowledges support from the Royal Swedish Academy of Sciences.


\end{document}